# Resampling-based batch reweighting for ComBat harmonization of MRI radiomic features in a multi-scanner glioma cohort

**Rifa Sajda Tasnim, M.Sc.[1], Yushuf Sharker, PhD[2], Md Forhad Hossain, PhD[3], Rakibur Rahman, PhD[4], Enamul Hoque Bhuiyan, PhD[5,6*]**

[1]Department of Biomedical Engineering, University of Toledo, Ohio, USA

[2]Center for Data Research and Analytics, Bethesda, MD, USA

[3]Department of Physics, University of Virginia, VA, USA

[4]Spiced Academy, Ritterstraße 12-14, 10969 Berlin, Germany

[5]Center for Magnetic Resonance Research, Department of Radiology, University of Illinois Chicago, Chicago, IL 60612, USA

[6]AI.Health4All Center, College of Medicine, University of Illinois Chicago, Chicago, IL 60608, USA

*** Correspondence:**
**Corresponding Author**
Enamul Hoque Bhuiyan, Center for Magnetic Resonance Research & AI.Health4All Center, College of Medicine, University of Illinois Chicago, IL 60612, USA. **Phone**: +1-312-413-1286, **Fax: +1-**312-355-1635**, email**: bhuiyan@uic.edu

**Running Title:**

Batch-Reweighted ComBat in Glioma

**Funding**

No external funding was received

**Abstract**

Radiomic features derived from magnetic resonance imaging (MRI) are sensitive to scanners; therefore, they require harmonization across scanners. In a radiomic feature dataset, some batches may be severely underrepresented compared to the others e.g., the Erasmus Glioma Database, which may provide harmonized features biased towards the larger batches in the dataset. We investigated whether reweighting underrepresented scanner batches before empirical-Bayes ComBat estimation improves feature harmonization and reduces residual scanner association while retaining grade-related information. In this retrospective study, the Erasmus Glioma Database included 733 eligible patients acquired at a single center across nine vendor-by-field-strength configurations. Structural MRI acquired at 1 T, 1.5 T, and 3 T; four structural MRI sequences were analyzed. From four structural MRI sequences, 428 features were computed using an IBSI-compliant PyRadiomics implementation. The sample sizes of some scanner batches were very small compared to others, which may produce unequal precision and shrinkage of batch-effect estimates. We proposed a data augmentation approach to overcome the limitation of ComBat. We provided exploratory evidence, along with confirmatory Kruskal-Wallis tests with a Benjamini-Hochberg correction to assess the proposed method. Before harmonization, 392/428 features (91.6%) differed significantly across scanner batches (median eta-squared=0.463). Standard ComBat reduced this to 13/428 (3.0%), whereas batch-augmented ComBat reduced it to 8/428 (1.9%); the median residual effect size was <0.001 for both harmonized conditions. Batch augmentation appeared to be a useful strategy to improve the performance of ComBat for feature harmonization under severe batch-size imbalance.

## 1 Introduction

Magnetic resonance imaging (MRI) is the primary modality for non-invasive characterization of brain tumors, providing essential information for diagnosis, treatment planning, and longitudinal monitoring (1–3). Unlike CT Hounsfield units, traditional structural MRI lacks a standard absolute intensity scale. As a result, measured intensities and derived texture features can differ based on scanner hardware, field strength, pulse-sequence implementation, reconstruction, and preprocessing (4–6). This variability poses a fundamental challenge for quantitative imaging analyses that rely on consistent signal properties across datasets.

Radiomics refers to the high-throughput extraction of quantitative features from medical images, including descriptors of shape, intensity distribution, and texture (7–9). These features have shown promise as non-invasive biomarkers for tumor grading, molecular mutation characterization, and outcome prediction in neuro-oncology (10–15). However, radiomic features are well recognized to be sensitive to differences in MRI acquisition parameters (16–22), which can introduce scanner-dependent variance that confounds biological signals and hampers the development of generalizable predictive models. Such variability makes it harder to combine data from different scanners and institutions (23,24). It can also lower the reproducibility and portability of radiomic models (25,26).

Several strategies have been proposed to reduce scanner-induced variability. At the image level, standardized acquisition protocols can reduce inter-site differences (27), though enforcing uniform protocols across institutions remains impractical for retrospective data. Image preprocessing pipelines, including bias field correction, voxel resampling, skull stripping, and intensity normalization, offer a second line of defence and have shown promising results in radiomic studies (28,29). Deep-learning harmonization methods have emerged, but they usually need large and representative training data. They may also need paired or domain-matched acquisitions based on the specific approach (30). At the feature level, the ComBat method, originally developed for batch-effect correction in genomic microarray data (31) and subsequently adapted for neuroimaging (32) and radiomic applications (33–38), has emerged as a widely used harmonization approach. ComBat models each feature as the sum of a grand mean, covariate effects, and batch-specific location and scale parameters, which are estimated using an empirical Bayes (EB) framework that borrows information across features to stabilize estimates for small batches.

Despite its effectiveness, ComBat has several practical limitations. First, it requires a statistically representative sample from each batch for stable EB estimation; the minimum sample size has been suggested as five per batch (32,39), though the quality of estimates degrades with smaller or highly imbalanced batches. Second, the definition of "batch" is not always straightforward: the same center may use multiple scanners, and the same scanner model may be operated with different protocols (36). Third, as Carré et al. (36) have emphasized, improper application of ComBat in machine learning pipelines—such as fitting on pooled train-test data—can introduce data leakage (40). Recent extensions, including AutoComBat (36), Nested ComBat (41), and GMM-based approaches (42), have addressed some of these limitations, but the problem of batch-size imbalance in EB estimation has received comparatively little attention. Empirical-Bayes shrinkage can stabilize batch-specific estimates when sample sizes are limited. However, severe imbalance in batch sizes changes the relative precision of batch-specific estimates and the extent of shrinkage toward the pooled prior. The effect of such imbalance on radiomic harmonization has received less systematic evaluation.

The question of how to define batch labels is a recognized challenge in multi-center radiomics (35,51). Da-Ano et al. (34) proposed unsupervised hierarchical clustering of radiomic features themselves to assign batch labels, while Carré et al. (36) developed AutoComBat, which uses DICOM metadata and image quality metrics as inputs for constrained clustering. In the present work, we adopted a simpler approach: defining batches as the combination of vendor and field strength. This choice was motivated by the availability of reliable vendor and field-strength metadata in our dataset and the observation that these two factors capture the dominant sources of scanner-induced variability (19,52). Future work could investigate whether finer-grained batch definitions based on DICOM metadata or image quality metrics, as proposed by Carré et al. (36), would further improve harmonization.

In retrospective multi-vendor datasets, batch-size imbalance is the norm rather than the exception. For example, a dataset pooled from multiple institutions may contain hundreds of patients scanned on a common Siemens 1.5T platform but only 20 patients from a Philips 3T scanner. For small batches, the posterior batch parameters shrink more toward the cross-feature prior than parameters from larger batches. Whether reweighting can reduce this imbalance and improve estimation without distorting biological variation is an empirical question.

In this work, we propose a simple batch-reweight strategy to address this imbalance. Before ComBat estimation, underrepresented batches are resampled with replacement to approximate the size of the largest batch, thereby increasing the effective weight assigned to underrepresented batches during ComBat parameter estimation. The augmented samples are retained only during the ComBat fitting step and are removed from the final harmonized dataset. We evaluated whether this strategy (i) reduces leftover scanner-related variation compared to standard ComBat, and (ii) maintains biologically related variation.

## 2 Materials and Methods

This study used the publicly available Erasmus Glioma Database (EGD), which was collected in accordance with the 1964 Helsinki Declaration and its later amendments. The original study was approved by the institutional review board of Erasmus MC, Rotterdam, The Netherlands. No additional ethical approval was required for this secondary analysis of de-identified, publicly available data.

The study workflow included selecting the cohort, preprocessing images, extracting radiomic features, reweighting scanner batches, harmonizing with ComBat, and systematically assessing residual scanner effects while preserving biologically relevant variation. We then conducted an exploratory analysis to classify tumor grades and evaluate their predictive usefulness. The complete analytical framework, including cohort preparation, harmonization strategies, and evaluation, is summarized in Figure 1.

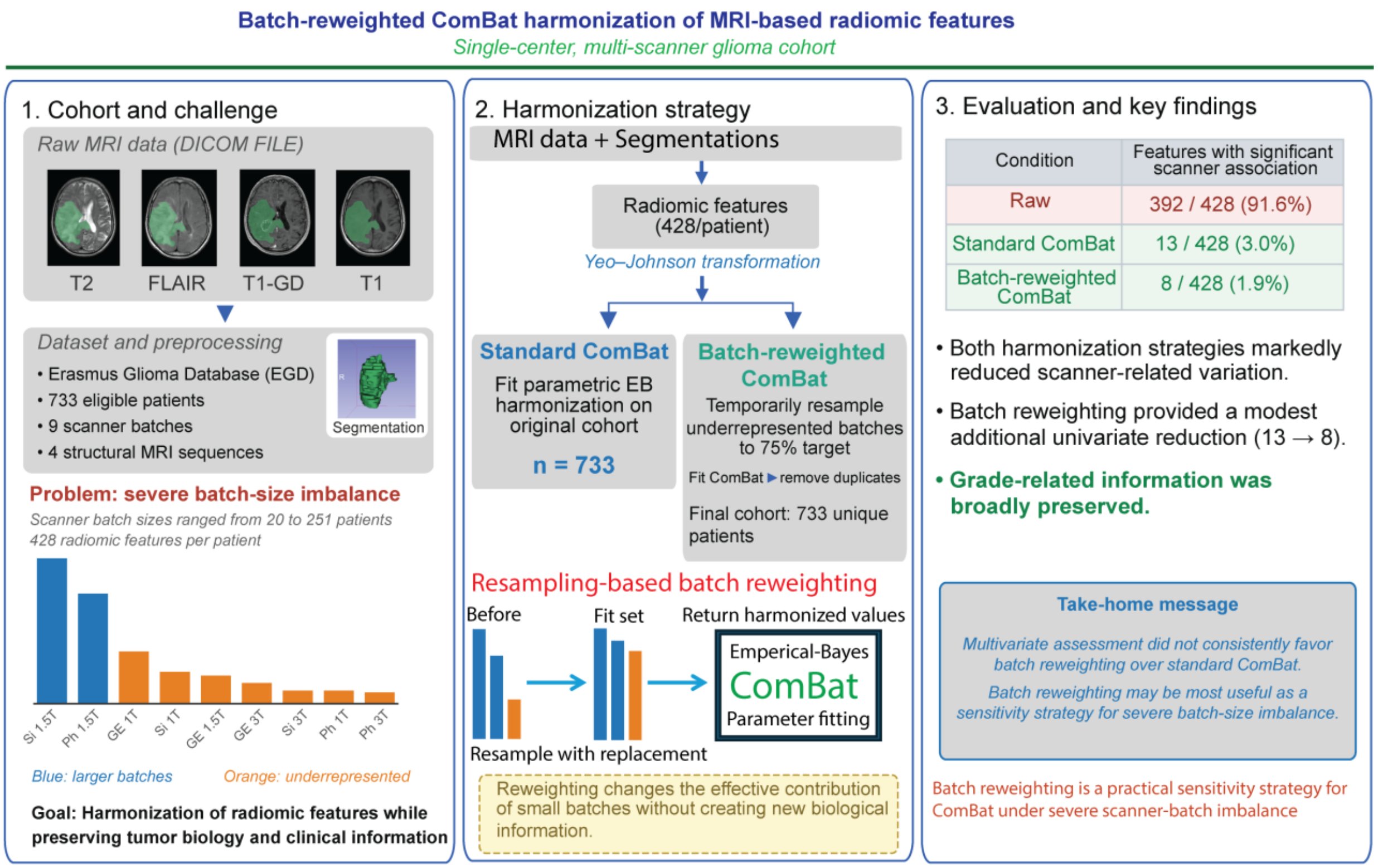

## 2.1 Study Cohort

Data were obtained from the Erasmus Glioma Database (EGD) (43), a publicly available dataset containing structural MRI scans, genetic and histological labels, and tumor segmentations of 774 patients with glioma treated at a single tertiary center (Erasmus MC, Rotterdam, The Netherlands) between 2008 and 2018. Despite originating from a single institution, the dataset exhibits substantial scanner heterogeneity, as imaging was performed on scanners from four vendors (Siemens, Philips, GE, Toshiba) at field strengths of 0.5T, 1T, 1.5T, and 3T. For each patient, T1w, contrast-enhanced T1w (T1w-ce or T1w-gd), T2w, and FLAIR MRI sequences were available as NIfTI files: pre-contrast T1-weighted (T1w), post-contrast T1-weighted (T1w-gd), T2-weighted (T2w), and T2-weighted FLAIR. The EGD distribution provides images registered to the MNI152 atlas at 1×1×1 $mm^3$ resolution. The whole tumor region was segmented by the EGD distributor. Radiomic features were extracted from the NIfTI files using 3D Slicer (44) (version 5.6.2) with the integrated PyRadiomics extension (45).

## 2.2 MRI and Radiomic Feature Extraction

Radiomic feature extraction. We extracted radiomic features using 3D Slicer (version 5.6.2) along with the SlicerRadiomics/PyRadiomics extension. The extraction took place in three dimensions, using predefined tumor segmentation masks. We only used unfiltered preprocessed MRI images for calculation; we did not include wavelet- or Laplacian-of-Gaussian-filtered images. For gray-level discretization, we used a fixed bin width of 25. We disabled PyRadiomics intensity normalization since the MRI volumes had already been preprocessed with the Erasmus Glioma Database. No additional voxel resampling occurred during radiomic extraction because the images were already spatially normalized to an isotropic voxel size of 1×1×1 $mm^3$. This way, we maintained the native preprocessed image geometry during feature calculation. We calculated three-dimensional texture features with force2D set to False, using a voxel distance of 1. The voxelArrayShift parameter was set to 0. The minimum required ROI dimensionality was 2, and masks needed to have valid spatial overlap with the image volume.

A total of 107 features were extracted per sequence, yielding 428 features per subject. Features comprised 14 shape descriptors, 18 first-order histogram statistics, and 75 second-order texture features derived from the Gray-Level Co-occurrence Matrix (GLCM), Grey Level Dependence Matrix (GLDM), Gray Level Run Length Matrix (GLRLM), Gray Level Size Zone Matrix (GLSZM), and Neighborhood Gray Tone Difference Matrix (NGTDM), compliant with the Image Biomarker Standardization Initiative (IBSI) (46).

2.3 Batch Definition and Exclusions

The scanner batch was defined as the combination of MRI manufacturer and field strength. Of 774 patients, 743 had complete radiomic feature data. Ten subjects were excluded because their scanner metadata did not allow reliable assignment to the predefined vendor, field strength categories. This included Toshiba systems, 0.5-T acquisitions, and cases with missing field-strength information. This left 733 patients across nine scanner batches, with batch sizes ranging from 20 to 251. No valid scanner configuration was excluded only due to small sample size.

2.4 Feature Normalization

Before harmonization, all radiomic features were transformed using the Yeo–Johnson power transform (47) to approximate normality within each batch, as recommended for radiomic feature preprocessing (48). Unlike the log transform, the Yeo–Johnson transformation handles features with negative values (e.g., ClusterShade, Skewness) without requiring arbitrary shifts. For each feature, a single transformation parameter was estimated from the analysis cohort and applied identically across scanner batches. Shape features, derived from the segmentation mask geometry and near identical across MRI sequences (Pearson $r > 0.97$), were harmonized once using T2w as the canonical sequence, and the corrected values were propagated to the remaining three sequences.

2.5 ComBat Harmonization with Empirical Bayes Shrinkage

Batch effects were corrected using the ComBat algorithm with parametric empirical Bayes shrinkage, following Johnson et al. (Johnson, 2007). The model decomposes each feature value $Y_{ijf}$ (feature f, sample j, batch i) as:

$$Y_{ijf} = \alpha_f + X\beta_f + \gamma_{if} + \delta_{if}\varepsilon_{ijf}$$

where $\alpha_f$ is the overall feature mean, X is a matrix for the covariates of interest (age and sex), $\beta_f$ is the corresponding regression coefficient vector, $\gamma_{if}$ and $\delta_{if}$ represent the additive (location) and multiplicative (scale) batch effects, respectively, and $\varepsilon_{ijf}$ is the residual error. The procedure consists of six steps: (1) ordinary least-squares estimation of batch intercepts and covariate coefficients; (2) standardization to Z-scores using pooled variance; (3) raw estimation of per-batch location shifts ($\hat{\gamma}$) and scale effects ($\hat{\delta}^2$); (4) empirical Bayes shrinkage, in which a Normal prior is placed on $\gamma$ and an Inverse-Gamma prior on $\delta^2$ across features, with hyperparameters estimated by method of moments, yielding posterior estimates $\gamma^*$ and $\delta^{2*}$; (5) correction of standardized values using the shrunken parameters, followed by restoration to the original feature scale; and (6) restoration of covariate effects to preserve biological variation attributable to age and sex.

ComBat was implemented using a parametric empirical-Bayes location-and-scale model with age and sex as covariates. Standard ComBat was fitted to the original 733 patients, whereas batch-reweighted ComBat used temporary within-batch resampling to increase the effective contribution of underrepresented scanner groups during parameter estimation. The augmented matrix contained 1,763 rows representing the same 733 unique patients; duplicated rows were removed after harmonization and were not used in subsequent analyses.

## 2.6 Resampling-Based Batch Reweighting for Imbalanced Batch Sizes

We performed row-wise sampling with replacement within each batch and included the sampled data in the dataset used for parameter estimation. Sampling continued until each batch reached at least 75% of the size of the largest batch. Seven of nine batches required augmentation to reach the 75% $n_{target} = [0.75max] = [188.25] = 189$. The working dataset increased from 733 to 1,763 subjects. The two largest batches (Siemens×1.5T, n = 251; Philips×1.5T, n = 189) remained unmodified. Augmented rows were removed after ComBat estimation, and the final harmonized dataset contained the original 733 subjects.

The nine scanner batches exhibited a 12.5-fold size imbalance (range, 20-251). For each batch with nb < 0.75 nmax, patients were sampled with replacement until the batch reached 0.75 nmax=189 observations. Sampling was stratified by scanner batch, and exact patient rows, including biological covariates, were replicated. This operation should be interpreted as batch reweighting rather than the generation of new biological information. Standard ComBat was fitted using the original 733 patients as the comparator.

## 2.7 Binary Tumor Grade Prediction

To assess whether harmonization preserved grade-related information, an exploratory binary tumor-grade classification analysis was performed using the original class distribution. Tumors were categorized as low grade (WHO grades 1–2; n=181) or high grade (WHO grades 3–4; n=551); one patient with missing grade information was excluded, yielding n=732 evaluable subjects. Under the unbalanced conditions, the original class distribution was maintained, while under the balanced conditions, random oversampling was carried out within each training fold.

## 2.8 Handling Class Size Imbalance in Training Data

The binary classification task exhibited a 1:3.0 class imbalance (low grade n = 181 vs. high grade n = 551). To address this, we implemented the same resampling-based batch reweighting method for the subjects from the low-grade tumor class. We randomly sampled subjects with replacement and appended them to the training dataset to get low- and high-grade sample sizes balanced (n = 551 each), yielding a balanced training set of 1,102 subjects.

We evaluate six experimental conditions to compare the performance of raw features, general ComBat-harmonized features, and data-augmented ComBat-harmonized features to predict the tumor grade with and without augmentation. Comprise all combinations of three feature sets (original raw features, ComBat without resampling-based batch reweighting, ComBat with resampling-based batch reweighting) crossed with two class-balance strategies (with and without ROR). In the text, “original” refers to the raw Yeo–Johnson-transformed features before any ComBat harmonization.

## 2.9 Classification Model and Cross-Validation

We used logistic regression with L2 regularization (L-BFGS solver; maximum 5,000 iterations). Features were standardized within each training fold. We summarized performance by accuracy, precision, sensitivity, specificity, F1 score, and ROC AUC using fivefold stratified cross-validation. The scikit-learn library (49) v0.23 was used for all machine learning pipelines.

## 2.10 Statistical Analysis

We compared the mean, standard deviation, skewness, and kurtosis of each feature across individual batches and the combined dataset. Kernel-density plots were generated before and after harmonization. For each of the 428 features and each condition (unharmonized, standard ComBat, and batch-augmented ComBat), residual scanner dependence was tested using a Kruskal-Wallis test across the nine batches. Effect magnitude was summarized using eta-squared, $\eta^2 = (H-k+1)/(N-k)$, where H is the Kruskal-Wallis statistic, k is the number of batches, and N is the number of patients. P-values were adjusted across features within each condition using the Benjamini-Hochberg false-discovery-rate procedure; $q<0.05$ was considered statistically significant. Analyses emphasized both the number of significant features and the distribution of effect sizes because significance alone is sample-size-dependent.

## 3 Results

### 3.1 Study Cohort

The study cohort comprised 733 patients distributed across nine scanner batches. Batch size ranged from 20 (Philips 3 T) to 251 (Siemens 1.5 T). The mean age of the patients was 56.3 years (range, 19-86), and 63.7% of patients were male. Grades were available for 732 patients: 181 (24.7%) were low grade (WHO grades 1-2) and 551 (75.3%) were high grade (WHO grades 3-4). The batch age, sex, and grade distributions are similar to the overall distributions (Table 1).

**Table 1. Baseline cohort characteristics stratified by scanner batch. Values are presented as mean ± SD for continuous variables and n (%) for categorical variables. Batches with n ≥189 were not resampled.**

| Scanner batch | Patients, n (733) | Age (years) Mean ± SD | Male n (%) | Low-grade n (%) | High-grade n (%) |
|---|---|---|---|---|---|
| Siemens 1.5T | 251 | 57.3 ± 13.7 | 158 (62.9) | 64 (25.5) | 187 (74.5) |
| Philips 1.5T | 189 | 59.3 ± 13.7 | 109 (57.7) | 41 (21.7) | 148 (78.3) |
| GE 1T | 89 | 55.8 ± 14.2 | 63 (70.8) | 26 (29.2) | 63 (70.8) |
| Siemens 1T | 55 | 57.7 ± 16.2 | 37 (67.3) | 12 (21.8) | 43 (78.2) |
| GE 1.5T | 48 | 52.7 ± 14.5 | 31 (64.6) | 16 (33.3) | 32 (66.7) |
| GE 3T | 35 | 54.5 ± 14.7 | 21 (60.0) | 8 (22.9) | 27 (77.1) |
| Siemens 3T | 24 | 57.8 ± 12.7 | 16 (66.7) | 5 (20.8) | 19 (79.2) |
| Philips 1T | 22 | 53.8 ± 13.9 | 14 (63.6) | 5 (22.7) | 17 (77.3) |
| Philips 3T | 20 | 57.3 ± 15.8 | 12 (60.0) | 4 (20.0) | 16 (80.0) |
| Overall | 733 | 57.1 ± 14.10 | 62.9% | 181 | 551 |

**Abbreviations: SD, standard deviation; GE, General Electric. Replace the low-/high-grade definition with your study-specific grading system if different; 1 = missing grade. Report 461 (62.9%) for males. Clearly state whether grade percentages use 732 evaluable patients or all 733 participants.**

## 3.2 Batch-Effect Removal

Before harmonization, Kruskal–Wallis (KW) tests identified significant scanner-batch effects after Benjamini–Hochberg false discovery rate correction (BH-FDR $q < 0.05$). Among 428 radiomics features analyzed, 392 (91.6%) showed significant scanner-batch effects (BH-FDR $q < 0.05$). The corresponding median effect size across features was $\eta^2 = 0.463$. Batch effects were observed in 24 of 56 shape features (42.9%), 70 of 72 first-order features (97.2%), and 298 of 300 texture features (99.3%).

## 3.3 Harmonization Using Standard ComBat

Following standard ComBat harmonization, the number of features exhibiting significant batch effects decreased to 13 of 392 (3.0%), with a median $\eta^2 < 0.001$, corresponding to a 96.7% reduction in the number of features with statistically significant residual scanner association.

## 3.4 Harmonization Using Batch-Augmented ComBat

Following batch-augmented ComBat harmonization, only 8 of 428 features (1.9%) retained significant batch effects, again with a median $\eta^2 < 0.001$. This represents a 98.0% reduction compared to the original dataset. The reweighted method retained five fewer FDR-significant associations. The 12 radiomic features with the strongest scanner batch effects are summarized in Table 2.

**Table 2. Radiomic features exhibiting the strongest scanner-related batch effects before harmonization. For each feature, eta-squared ($\eta^2$) and Benjamini–Hochberg false-discovery-rate-adjusted q-values are reported before harmonization, after standard ComBat harmonization, and after batch-reweighted ComBat harmonization. Supplementary Tables S1, S2 & S3 provide feature-wise descriptive statistics (mean ± SD) for all nine scanner batches under each harmonization condition.**

| Features | Mean | Median | SD | Skewness | Kurtosis | $\eta^2$ Original | q Original | $\eta^2$ ComBat | q ComBat | $\eta^2$ Batch-Reweighted ComBat | q Batch-Reweighted ComBat |
|---|---|---|---|---|---|---|---|---|---|---|---|
| T2 Inverse Variance | 1.8160 | 2.0788 | 0.9078 | -0.3215 | -1.3154 | 0.618 | <0.001 | 0.014 | 0.228 | 0.012 | 0.432 |
| T1GD Root Mean Squared | 1.8160 | 1.8212 | 0.5406 | -0.3214 | 2.8122 | 0.590 | <0.001 | 0.006 | 0.828 | 0.005 | 0.938 |
| T1GD Mean | 1.7255 | 1.7310 | 0.5990 | -0.3475 | 2.8323 | 0.587 | <0.001 | 0.006 | 0.854 | 0.005 | 0.933 |
| T1GD Median | 7.1164 | 7.0668 | 1.1387 | 0.0291 | 2.3549 | 0.582 | <0.001 | 0.005 | 0.854 | 0.006 | 0.897 |
| T1GD 90Percentile | 7.1164 | 7.2249 | 0.6982 | -1.1887 | 2.5707 | 0.581 | <0.001 | 0.001 | 1.000 | <0.001 | 1.000 |
| FLAIR Inverse Variance | 7.0636 | 7.1718 | 0.7101 | -1.2655 | 3.2791 | 0.575 | <0.001 | 0.011 | 0.380 | 0.008 | 0.649 |
| T1 90Percentile | 7.3151 | 7.2701 | 1.2030 | 0.0345 | 2.1749 | 0.565 | <0.001 | <0.001 | 1.000 | <0.001 | 1.000 |
| T1 Root Mean Squared | 7.3151 | 7.4153 | 0.7415 | -1.1876 | 2.5606 | 0.553 | <0.001 | <0.001 | 1.000 | <0.001 | 1.000 |
| T1GD Maximum | 7.2594 | 7.3671 | 0.7485 | -1.2520 | 3.1850 | 0.552 | <0.001 | <0.001 | 1.000 | <0.001 | 1.000 |
| T1 Mean | 7.5758 | 7.5253 | 1.2835 | 0.0403 | 1.9816 | 0.550 | <0.001 | <0.001 | 1.000 | <0.001 | 1.000 |
| T1GD Large Area Low Gray Level Emphasis | 7.5758 | 7.6977 | 0.7967 | -1.1529 | 2.5090 | 0.550 | <0.001 | <0.001 | 1.000 | <0.001 | 1.000 |
| T2 Run Variance | 7.5124 | 7.6341 | 0.8002 | -1.1994 | 2.9878 | 0.546 | <0.001 | 0.007 | 0.701 | 0.008 | 0.649 |

**Note: For descriptive statistics, see Supplementary Tables S1, S2 & S3.**

Before harmonization, all features exhibited extremely strong scanner effects (q-values approaching zero; $\eta^2 = 0.55–0.62$). Both standard ComBat and Batch-Augmented ComBat effectively removed scanner-related variation. For features with residual scanner effects after standard ComBat, Batch-Augmented ComBat.  For the selected features, residual batch associations were generally comparable or smaller. For example, T2 Inverse Variance improved from $\eta^2 = 0.014$ to 0.012 and $q = 0.228$ to 0.432, while FLAIR Inverse Variance improved from $\eta^2 = 0.011$ to 0.008 and $q = 0.380$ to 0.649. For the remaining features, both methods reduced scanner effects to negligible levels ($\eta^2 < 0.01$; $q \geq 0.70$ or 1.00), with Batch-Augmented ComBat matching or modestly improving upon standard ComBat.

Fig. 2(a–l) shows the distributions of the same 12 radiomic features tabulated in Table 2. The original data exhibits clear scanner-specific differences, indicating substantial batch effects. Both harmonization methods markedly improve the alignment of scanner-specific distributions while preserving the overall feature distributions. Across the 12 features, Batch-Augmented ComBat achieves comparable or slightly better alignment than standard ComBat, particularly for features with residual scanner effects, consistent with the quantitative results in Table 2.

Complete per-batch descriptive statistics (mean ± SD) for all nine scanner batches are provided in Supplementary Tables S1, S2 & S3.

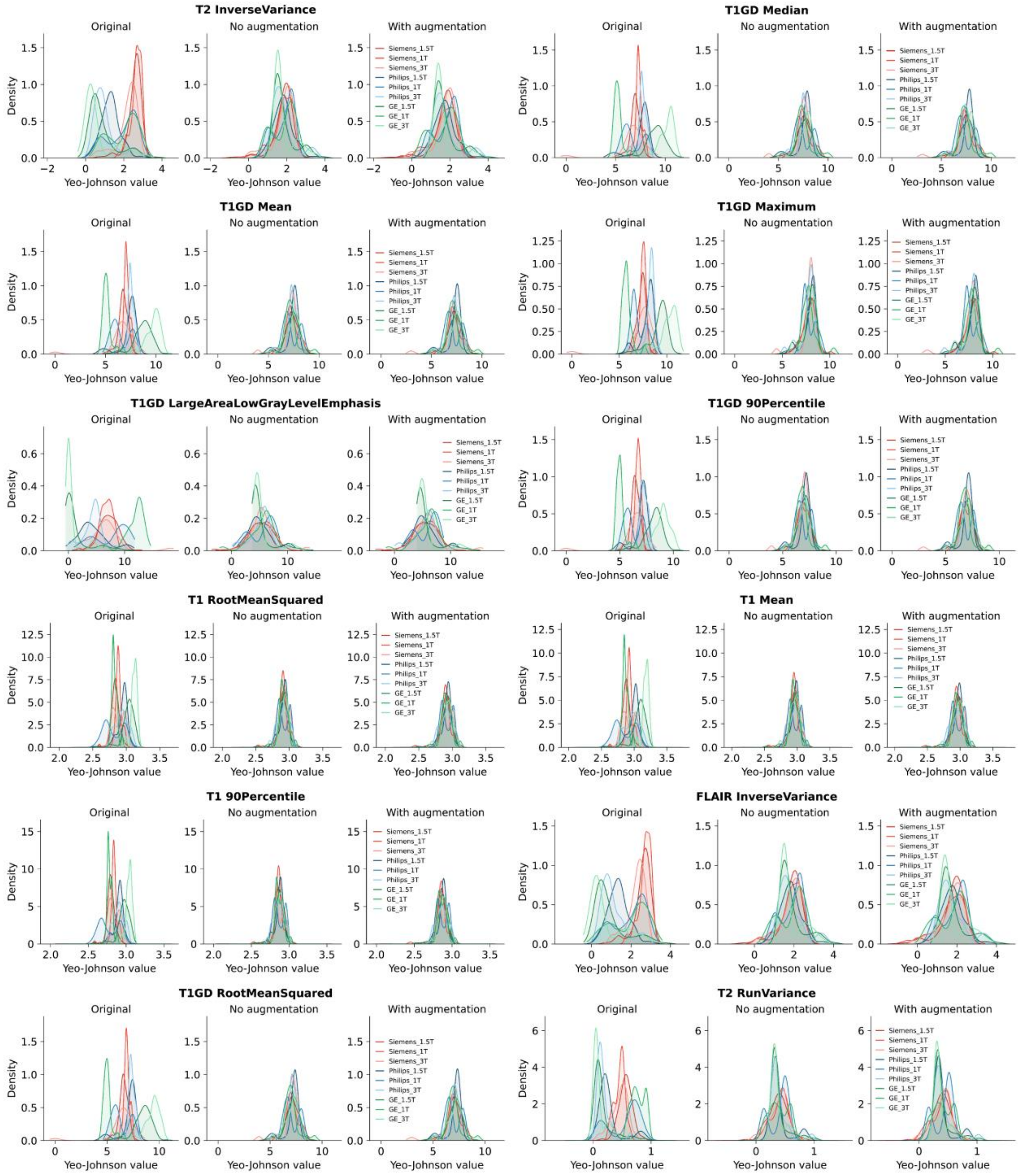

**Figure 2.** Three-way density comparisons for the 12 features with the highest pre-harmonization batch effects ($\eta^2$ range: 0.546–0.618). Each panel shows kernel density estimates per scanner batch under three conditions: original (Yeo–Johnson transformed raw features), ComBat without resampling-based batch reweighting (BA), and ComBat with BA. Each curve represents one of nine scanner batches (color-coded by vendor: red = Siemens, blue = Philips, green = GE). In the original condition, batch-specific distributions are widely separated; after ComBat with BA, distributions converge substantially. **Abbreviations:** BA, batch augmentation; FLAIR, fluid-attenuated inversion recovery; GE, General Electric; SD, standard deviation.

### 3.5 Exploratory Binary Grade Prediction

Without augmentation, the models favored the high-grade majority class, yielding sensitivity of 86.9-89.5% but specificity of 47.0-49.7%. After minority-class resampling, specificity increased to 82.2-84.8%, and AUC values ranged from 0.851 to 0.856 (Table 3).

**Table 3. Binary grade prediction performance across six experimental conditions.**

| Condition | Population (N) | Acc % | Perc % | Sens % | F1-score % | Spec % | Area under the curve (AUC) |
|---|---|---|---|---|---|---|---|
| Original | 732 | 79.0 | 83.7 | 89.5 | 86.5 | 47.0 | 0.763 |
| Original + Data Aug | 1102 | 79.9 | 81.3 | 77.5 | 79.4 | 82.2 | 0.854 |
| ComBat (no BR) | 732 | 77.0 | 83.3 | 86.9 | 85.1 | 47.0 | 0.745 |
| ComBat (no BR) + ROR | 1102 | 80.6 | 83.4 | 76.4 | 79.7 | 84.8 | 0.851 |
| ComBat (with BR) | 732 | 77.9 | 84.1 | 87.1 | 85.6 | 49.7 | 0.752 |
| ComBat (with BR) + ROR | 1102 | 80.9 | 83.2 | 77.3 | 80.2 | 84.4 | 0.856 |

**Abbreviations: Acc, accuracy; AUC, area under the receiver operating characteristic curve; BR, batch reweighting; Prec, precision; ROR, random oversampling with replacement; Sens, sensitivity; Spec, specificity. Original denotes raw Yeo–Johnson-transformed features.**

Differences among the three harmonization conditions were small. These estimates should not be interpreted as external clinical performance because they arise from internal cross-validation in one retrospective cohort.

Class balancing improved recognition of the minority class but reduced sensitivity for the high-grade class. The highest tabulated AUC (0.856) occurred for batch-augmented ComBat with grade balancing; however, the difference from the unharmonized balanced condition (AUC=0.854) was negligible, and no confidence intervals or paired statistical comparisons were available. Accordingly, the downstream analysis supports preservation, rather than improvement, of grade-related information.

Similar ROC curves within the balanced and unbalanced groups demonstrate a near overlap among harmonization variants, suggesting that the dominant change in the tabulated classification results was due to class balancing rather than ComBat (Fig. 3). A leakage-safe nested analysis with confidence intervals is required to quantify any incremental predictive effect of resampling-based batch reweighting.

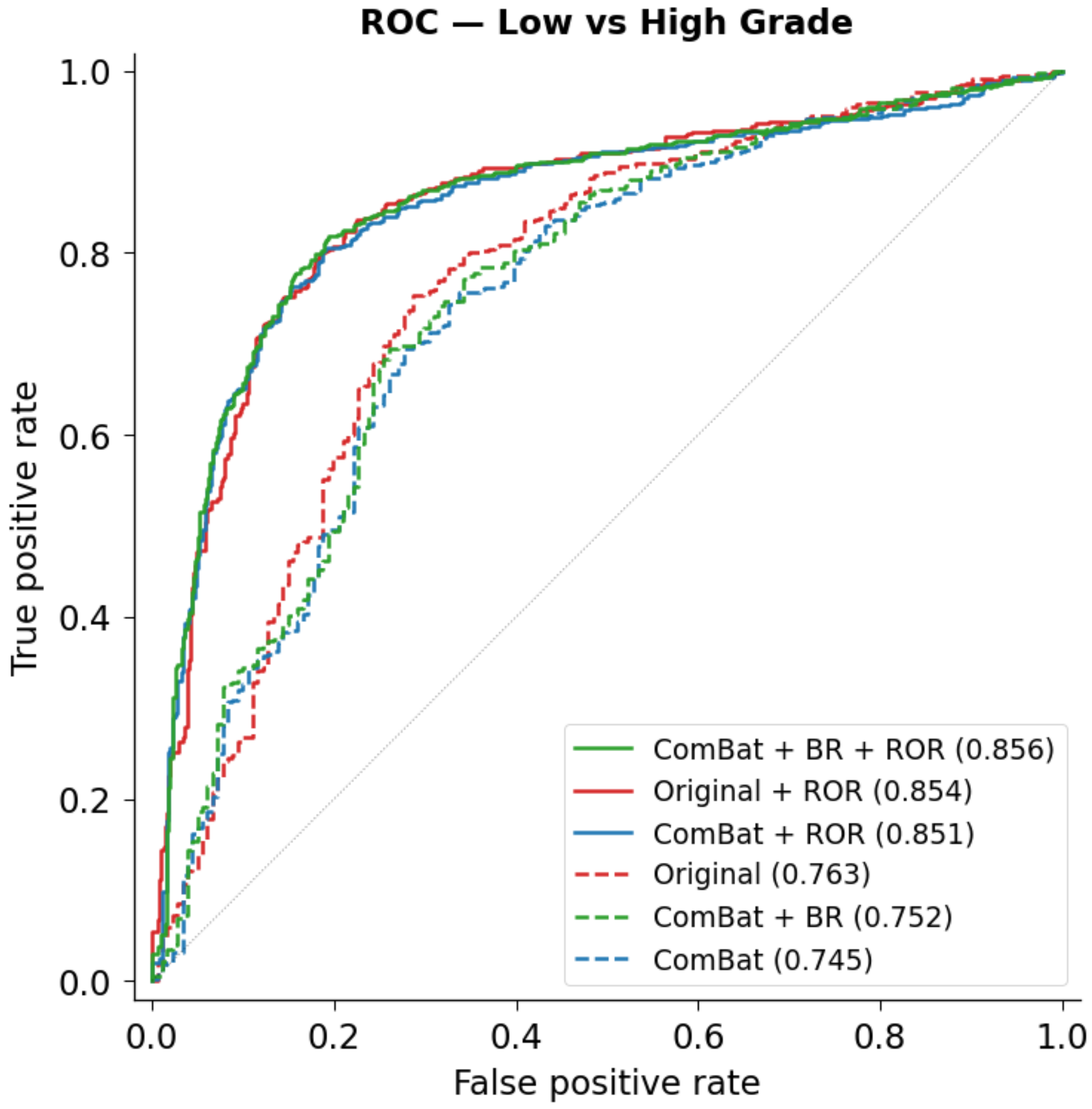

**Figure 3.** Receiver operating characteristic curves for exploratory binary low- versus high-grade classification under six combinations of feature harmonization and random oversampling. Curves are shown for original, standard ComBat, and batch-reweighted ComBat feature sets with and without random oversampling. Abbreviations: AUC, area under the curve; BA, batch augmentation; ComBat, combining batches; ROC, receiver operating characteristic; ROR, random oversampling with replacement.

Random oversampling of the minority class greatly improved the ability to discriminate; the ROC AUCs ranged from 0.745 to 0.763 without oversampling and from 0.851 to 0.856 with oversampling. The performance was very similar among the three feature sets that had been oversampled, the highest AUC being that of batch-reweighted ComBat (AUC = 0.856), then that of the original (AUC = 0.854) and finally that of the standard ComBat data (AUC = 0.851). The fact that the ROC curves were so closely overlapping shows that the harmonization had preserved the information related to grade but had not resulted in any significant improvement in discriminatory performance.

## 4 Discussion

Systematic differences in radiomic feature distributions that are attributable to the imaging “batch” (e.g., scanner, site, acquisition protocol), while attempting to preserve biological variation that is not related to the batch

In this study conducted at a single center using multiple scanners, the batch-reweighted ComBat outperforms ComBat and no harmonization in eliminating the systematic differences in radiomic feature distributions that are attributable to the imaging “batches” and preserving the biological variations that were not related to the batch. These results suggest that batch reweighting could be a useful approach for dealing with batch-size imbalances in radiomic features to be used in analysis.

With the proposed method, the number of features with notable scanner-batch association dropped by 98.0% after batch-reweighted ComBat. This builds on earlier findings across nine different vendor and field-strength configurations. Orlhac et al. (33) showed that ComBat combined with image preprocessing could eliminate statistical differences between radiomic features extracted from 1.5T and 3T images. Carré et al. (28) reported that ComBat applied directly to raw images achieved 68% and 79% harmonization strength augmentation for T1w-gd and T2w-flair images, respectively, and proposed that ComBat implicitly learns some sources of variability, thereby bypassing certain preprocessing steps. Da-Ano et al. (34) compared four versions of non-parametric ComBat and found that all reduced machine differences, with the reference-site version performing best (50). Our results, showing a 98% reduction in batch-affected features across nine batches, extend these findings to a larger number of scanner configurations and a more heterogeneous vendor–field-strength landscape.

Batch definition is a key factor in radiomic harmonization. In this study, we defined scanner batches based on the manufacturer and field strength since these details were consistently available across the group. This definition should be regarded as a stand-in for the acquisition domain, not as a way to identify a specific scanner. Scanners from the same manufacturer and with the same field strength can still vary in model, software version, coil configuration, reconstruction, and sequence settings. More detailed batch definitions that include this kind of metadata might further improve our understanding of acquisition-related variability, provided that we have enough complete metadata.

Tumor-grade classification was included mainly to assess how well biological signals were preserved, not as a clinical prediction experiment. The accuracy of classification was similar for the original, standard-ComBat, and batch-reweighted feature sets. This suggests that harmonization did not significantly reduce grade-related information. Since this analysis was done within a single retrospective cohort, these findings should be seen as supporting evidence for biological-signal preservation rather than for generalizing predictions externally.

Both harmonization approaches substantially reduced scanner-associated feature variation while preserving grade-related information. This distinction is important: harmonization should be judged primarily by removal of nuisance variation while preserving biological information, not by an expectation that it will necessarily increase within-cohort AUC.

The proposed procedure should be seen as resampling-based batch reweighting. Replication boosts the effective impact of smaller batches on ComBat parameter estimation and changes empirical-Bayes shrinkage, but it does not add independent information. Therefore, improvements in harmonization should not be viewed as increases in sample information. A natural extension would be a weighted or uncertainty-aware ComBat formulation that adjusts batch contributions directly, without treating replicated observations as independent samples. Standard ComBat effectively corrected scanner-related differences in radiomic data. Batch reweighting offered a slight extra benefit for some univariate features, especially when there was a large imbalance in batch sizes. Therefore, batch reweighting should be seen as a strategy for sensitivity rather than a guaranteed improvement over standard ComBat. It will be essential to validate this approach across different institutions and conduct simulations with known batch effects to identify when reweighting delivers consistent benefits.

## 5 Limitations

This study has several limitations. First, the analysis used a single-center cohort, which limits how well we can apply the findings to larger multi-institutional settings. Second, scanner batches were defined only by vendor and field strength, so they did not consider protocol, reconstruction, coil, or time variations. Third, there was no independent or traveling-subject dataset to confirm the removal of scanner effects. Finally, resampling-based reweighting results must be validated by multiple datasets.

The batch-reweighting method may not work well with very small batch sizes. To apply it effectively, each batch should contain enough samples to provide sufficient variability and ensure that the sampled data contributes meaningfully to the dataset. A minimal batch size of more than 10 may be considered as a rule of thumb. The approach involves resampling, so results may vary across runs. Using a fixed random seed is strongly recommended to ensure reproducibility.

## 6 Conclusion

In this highly imbalanced, single-center, multi-scanner glioma cohort, batch reweighting based on resampling reduced residual scanner-related radiomic features. Batch-reweighted ComBat outperformed standard ComBat. Although both methods greatly improved radiomics feature harmonization, they still retain discriminatory information related to tumor grade. The results indicate that batch reweighting could be a promising extension of ComBat in cases where the acquisition groups are greatly unbalanced; nevertheless, its reproducibility and predictive power need to be confirmed by repeated resampling, leakage-safe validation, and independent multicenter evaluation. These findings motivate further evaluation of batch-size reweighting through repeated resampling, ground-truth simulations, and independent multicenter validation before routine use

## Acknowledgments

The authors acknowledge the Erasmus Glioma Database (EGD) team for making their dataset publicly available. We used Grammarly for grammatical corrections, and OpenAI Codex (GPT-5, OpenAI, 2026) was used to assist with manuscript formatting. The authors reviewed the resulting document and took responsibility for its content.

**Author Contributions**

R.S.T.: data curation, methodology, investigation, formal analysis, visualization, and writing - original draft. Y.S.: statistical methodology, statistical analysis, visualization, validation, and writing - review and editing. M.F.H.: formal analysis, visualization, validation, writing - review and editing. R.R.: visualization, validation, machine learning, and writing - review and editing. E.H.B.: conceptualization, investigation, visualization, validation, supervision, project administration, resources, methodology, and writing - review and editing, and writing - original draft.

**Declaration of Competing Interests**

The authors declare no competing interests.

**Conflict of Interest**

The authors declare no Conflict of interest

**Data Availability Statement**

The EGD imaging dataset is available at https://xnat.bmia.nl/data/archive/projects/egd and is described at https://doi.org/10.1016/j.dib.2021.107191. Code repository URL and DOI to be shared upon acceptance.

**Supplementary Material**

Supplementary Table S1, S2 & S3